\documentstyle[11pt,paspconf,epsfig]{article}

\def\gtsim{\mathrel{\hbox{\rlap{\hbox{\lower4pt\hbox{$\sim$}}}\hbox{$>$}}}}
\newcommand{\ea}{et~al.\,}
\newcommand{\kms}{km~s$^{-1}$}
\newcommand{\msun}{M$_{\odot}$}
 
\begin{document}
 
\title{Horizontal-Branch Stars:\\ 
      Their nature and their absolute magnitude}

\author{Klaas S. de Boer}

\affil{Sternwarte, University of Bonn,\\
Auf dem H\"ugel 71, D-53121 Bonn, Germany}

\begin{abstract}
Horizontal-branch stars play an important r\^{o}le in defining the 
absolute distance scale in the universe. 
For that one has to know $M_V$ at relevant $B-V$ with well defined [M/H] 
for each type of horizontal-branch star. 
The nature of the horizontal-branch stars is reviewed with emphasis 
on all the open questions regarding evolution and structure, 
questions which show that these stars are not the easiest objects 
in the distance scale discussions. 

Using the HB stars with the best Hipparcos parallaxes, 
the HB  at $(B-V)_0 = 0.20$ and  [Fe/H] = $-1.5$ has $M_V = +0.71$.
Assuming the [Fe/H] to $M_V$ relation for RR Lyrae stars to be valid for 
HB stars too, one finds $M_V \simeq 1.00$ for [Fe/H] = 0.
\end{abstract}

\section{The stars}

Stars of the horizontal branch (HB) are post red giant (RG) stars 
which burn helium in their core. 
The mass of the He in the core is thought to be $\simeq 0.5$ \msun. 
Along the HB in the colour-magnitude diagram (CMD) 
one recognizes (from hot to cool) the types 
sdOB, sdB, HBB, HBA, RR Lyr, and RHB 
(for definitions see nomenclature review by de Boer \ea 1998). 
Depending on the initial parameters of the progenitor main sequence star 
(initial mass, metal content, etc.) as well as 
on aspects of the RG mass loss phase, 
the star will retain a hydrogen shell of a certain mass. 
If sufficient hydrogen remained there is also a hydrogen burning shell. 

The HB stars have a low to normal atmospheric metal content. 
Naturally, the colour of the star is determined by 
the atmospheric structure ($T_{\rm eff}$, $\log g$, [M/H]). 
HB stars show no or only very slow rotation (Peterson 1983). 

A large variety of models for HB stars have been calculated 
(e.g., Sweigart \& Gross 1976; Sweigart 1987; Dorman 1992) all indicating 
that total mass is positively correlated with luminosity and redness. 
At the very blue end of the HB the mass of the stars is $\simeq 0.5$ \msun\ 
(vanishingly thin H shell, no shell burning), 
at the red end it may amount to $\simeq 1.0$ \msun\ or more 
(thick shell and well established H shell burning). 

In metal poor globular clusters the HB is populated at the blue end, 
while in more metal rich globular clusters the HB stars group together 
in a small colour range at the red end, there forming a `red clump' 
seemingly associated with the red giant branch. 
However, this picture is rather simple and observations show that 
further parameters are required 
to also explain those globular clusters deviating from that scheme. 

\section{Evolutionary history and the HB stars as a group}

What is the evolutionary history of stars now being in the HB phase? 
With which mass did these stars start on the main sequence?
Models indicate that all stars initially having $0.8 \leq M \leq 3$ \msun\ 
do become HB-like stars. 
This range immediately implies that the HB stars intrinsically 
span a wide range in age and thus probably also in metal content. 
A star starting with 3 \msun\ will evolve into an HB star 
within $\simeq 10^9$~y. 
It must thus be regarded as `young' and 
have formed in the disk from material with most likely `solar' composition. 
Stars starting with 0.8 \msun\ will need $\geq 10^{10}$ y 
to become HB like, can thus be regarded as `old' objects and therefore 
must have formed from material substantial poorer in metals than the Sun. 

Since it is virtually impossible to determine the age of an individual 
{\it field\,} star with any certainty, 
one cannot discriminate a young from an old field HB star, 
except perhaps when assuming some age-metallicity relation. 
We must acknowledge that the field HB stars we have access to 
are a mix of young and old stars. 
With a more or less constant star formation rate in the Milky Way 
the HB star group forms a continuum in age 
but the older ones are likely to dominate in number. 

\section{HB star metal content and its consequences}

The determination of the metal content in the atmospheres of stars 
is with present day technology in principle not very difficult. 
Spectroscopic investigations show that 
the sdB stars have near normal Si with somewhat low He (Heber \ea 1984). 
The HBA stars, among which are the long known classical field HB stars, 
range in metal content from solar to [M/H] $\simeq -2.0$ 
using photometric methods (e.g.\,Gray \ea 1996), 
while spectroscopic determinations (naturally for just a few stars)
indicate generally low metal abundances, such as small [Fe/H], 
in part as low as in metal poor globular cluster giants
(Kodaira \& Philip 1984; Adelman \& Philip 1994). 
RR Lyr stars are known to spread over a large range in metal content 
(see e.g. Lambert \ea 1996, Layden 1994) 
with a predominance of metal poor ones. 
The RHB stars have not been studied intensively yet. 

Many metal abundance values are based on the measurement of some line index, 
like the Ca index, or are based on the $\Delta S$ method, 
giving a metallicity index to be denoted with [M/H]. 
The index is then calibrated with the help of spectroscopic [Fe/H] values 
(Lambert \ea 1996, Layden 1994). 
Such studies lead, for RR Lyrae stars, to a relation of $M_V$ versus [Fe/H] 
in which $M_V$ is taken from Baade-Wesselink methods (see Fernley 1994). 
However, one has to acknowledge that such metallicity indices are 
often not very accurate. 
When metallicities are given\footnote{Only when Fe lines are seen 
and analysed indeed, the use of [Fe/H] is justified} 
one always should trace their origin 
(are they photometric or spectroscopic\footnote{Also spectroscopic abundances 
may have substantial uncertainties; when the $T_{\rm eff}$ choosen for the 
analysis is different from reality, the line excitation model is off too, 
leading to errors in metal abundances easily of the order of 0.1 dex} values?) 
and one should be wary of too many decimals which often come only 
from a numerical transformation using rather uncertain input values. 

Also for globular clusters the photometric indices are problematic. 
This is of relevance here, since spectroscopic abundance determinations 
are available only for a limited number of clusters. 
Also the $Q_{39}$-index gives ambiguous results (de Boer 1988). 
This can be seen in Fig.\,12 of Zinn \& West (1984), 
in which the discrepancy between the metallicity 
from the $Q_{39}$-index and that from spectroscopy 
is seen to range from $\Delta$[Fe/H] of $-0.3$ to $+0.3$, 
depending on the metallicity and type of cluster. 
The discrepancies for clusters with blue HBs show a clear trend. 
The reason is, of course, that the photometry uses the entire cluster 
in which the various star types 
(red giants, main-sequence stars, red and blue HB stars) 
contribute according to their brightness and their number in the cluster 
(in particular a blue or a red HB),
thus mixing stars of widely differing temperature and gravity into 
one photometric index measurement. 
The widely used Zinn \& West (1984) metallicity scale 
clearly is of limited accuracy (de Boer 1988). 

Can one use the abundance values to estimate the age of each star?
Unfortunately, these abundances are not necessarily giving 
the intrinsic metal content of the stars. 
HB-like stars have a rather stable atmosphere. 
This means that in particular in conditions of high surface gravity, 
such as in sdOB and sdB stars, 
the heavy elements can diffuse downward and sink out of the atmosphere 
so that the star looks more metal poor than it is intrinsically. 
The effect was first explained for White Dwarfs by Michaud \ea (1984). 
In the RHB stars on the other hand, with low surface gravity, 
the radiation field may levitate the heavy elements (see Cassisi \ea 1997), 
so that their atmospheres may look richer in metals than the star really is. 
Furthermore, due to convection during the RGB phase He may be mixed 
into the surface layers of the star, possibly producing He-rich HB stars. 
For such He enriched HB star atmospheres, Sweigart (1997) speculates 
that the mixing in of He results in a bluer HB morphology 
(see also Sect.\,9), possibly an enlarged RR Lyrae period shift, 
and lower surface gravities in blue HB stars. 

According to model calculations, 
the absolute magnitude $M_V$ of HB-like stars depends on [M/H] in the sense 
that metal poor stars are brighter than metal rich stars. 
This is caused by an intricate interplay between internal opacity 
and the luminosity of the H-burning shell (Dorman 1992) 
and has no detailed relation with the metal content in the photosphere 
(see further Sect.\,9). 
 
Summarizing, the determination of the abundance of the elements 
leads to knowledge about the composition of the stellar surface. 
For inhomogeneous envelopes these atmospheric abundances are not related 
with the overall [M/H] of the star. 
The stellar structure models must take these inhomogeneities into account, 
lest they predict brightnesses and colours which do not conform with reality. 
Vice-versa, interpretation of the observed colour and brightness with 
current models may lead to faulty atmospheric and structural parameters. 

\section{Are `red HB stars' really red HB stars?}

The location in the CMD where red HB stars appear 
contains numerous other kinds of stars. 
Therefore, knowing if a red {\it field\,} star is a RHB star is not easy, 
because the appearance of a star with colour and magnitude like that 
of a red HB star is not sufficient proof. 

The red part of the HB crosses the red giant branch (RGB). 
This is well known from the study of the redder (less metal poor) 
globular clusters, where in some cases the RHB stars group 
together into the `red clump'. 
For the field stars the above must be true, too. 

It has become clear from models that the evolution on the RG branch has 
temporal irregularities (depending on which parameter one considers). 
This is related to the passage of the H-burning shell 
through the chemical discontinuity 
left by the convective envelope during the first dredge up phase 
(see e.g.\,Cassisi \ea 1997). 
It results in a so called RGB bump, a location where the RGB is 
(relatively) overpopulated (at $M_V \simeq +0.5$ mag). 

Due to the above coincidence, 
the RHB and the RGB bump appear to lie almost at the same $L$ and $T$ 
(depending on the age and the metallicity of the star group) 
and thus at virtually the same position in the CMD. 
This implies that it is for field stars almost impossible to discriminate 
between a RG star and a RHB star. 

A further problem is that blue loop stars 
(more massive stars which land in a more luminous core He burning phase) 
appear in colour magnitude diagrams very close to where the HB is. 
At the reddest and faintest point, these stars lie at $M_V = -0.5$ mag. 

The mentioned evolutionary states lead in complicated ways to 
an enhancement of stars in the CMD, as nicely illustrated by Gallart (1998). 
A summary of evolutionary details is given by Girardi \ea (1998). 
Clearly, stars in the RHB domain of a colour magnitude diagram for 
{\it field stars\,} cannot simply be taken to be RHB stars indeed. 
 
\section{Gaps and other structure on the HB}

Newell (1970) presented the first evidence for an irregular distribution 
of stars along the horizontal branch. 
In fact, he showed that there is evidence for two `gaps', 
regions on the HB with reduced numbers (or even devoid) of stars. 
This has been corroborated since then many times in newer 
and more extensive data sets, 
such as those collected for the sdB studies in Bonn (Aguilar S\'{a}nchez 1998).
The gaps are also present in the $T_{\rm eff}$ versus $\log g$ diagram. 
Globular cluster HB's show gaps as well. 
A recent detailed study is that by Ferraro \ea (1998). 
The cause for the gaps is largely unclear and 
several possibilities are being investigated. 

One possibility is that, in globular clusters, 
HB stars are present with two different mass ranges. 
Catelan \ea (1998) have simulated globular cluster HB distributions 
using both unimodal and bimodal mass distributions. 
The gaps do seem to be present but not with great significance. 
Even in the unimodal mass distribution sparse regions can occur. 

Another possibility for the occurrence of gaps is that the HB stars behave 
such that certain ranges of $T_{\rm eff}$ (or of $B-V$) are not present. 
I speculate that a gap may result from a small discontinuity 
in the burning in the hydrogen shell. 
If the energy production in a marginally burning shell is 
temporarily decreased (e.g.\,by stochastic fluctuations) 
the ensuing drop in temperature may be sufficient for 
the burning not to come up to the original level any more, 
and the hydrogen burning may extinguish altogether. 
The atmosphere then would become more compact and the colour of the star 
more blue than before. 
Thus at that HB location a gap in the smooth distribution along the HB 
may be created. 

A further possibility for the gaps to emerge is when the HB 
is populated by stars having different evolutionary origins. 
If the genesis of HB stars follows different routes, 
such as one kind directly from the RGB, 
the other through some binary star evolution (see e.g. Iben \& Tutukov 1985), 
the two routes may lead to different final masses 
and possibly to preferred locations on the HB. 

HB stars do seem to deviate in colour from expected values 
in certain cases, too. 
Grundahl \ea (1998) noted from Str\"omgren photometry that 
the shape of the distribution of the stars on the HB in the $u-y$ colour 
in the globular clusters M~13, NGC~288 and 6752, 
deviates from that of theoretical models. 
Either the stars are brighter than expected over a part of the blue HB, 
or they are bluer than expected. 
They speculate that there are two populations even for HB stars 
within a globular cluster. 

New HST data on a few metal rich globular clusters have shown that 
these clusters do have a population of blue HB stars, 
in contrast to established beliefs for metal rich clusters (Rich \ea 1997). 
Not only that, the HB's appear to become brighter toward the blue, 
in stark contrast to what has thusfar been found. 

Summarising, the shape of the observed HBs has large variations which 
the standard models cannot explain. 
What consequences have the gaps, the in-cluster differences, 
and the different possibe origins of the HB stars 
for our understanding of field HB stars?

\section{HB star mass and absolute magnitudes}
 
Models indicate that the mass of HB stars runs from 0.50 \msun\ at the 
blue end to $\simeq 1.0$ \msun\ at the red end of the HB. 
The luminosity runs from $\log L = 1.2$ to 1.7 L$_{\odot}$, respectively. 
Both mass and luminosity can only be found for stars with known distance. 

In spectroscopic studies, which result in $T_{\rm eff}$, $\log g$, and 
the apparent luminosity $l$, masses of the HB stars in globular clusters 
(distance known) have been determined.
It was found that 
the HBB stars in NGC~6397 (de Boer \ea 1995) 
and the sdB stars in M~15 (Moehler \ea 1995) 
have a mass of $\simeq 0.4$ \msun, 
much lower than the canonical value of 0.5 to 0.6 \msun\ for such stars. 
Both NGC~6397 and M~15 have the very low metallicity ([M/H] $\simeq -2$ dex). 

Masses for RR Lyrae stars can be calculated from the relations of 
period, luminosity and mass. 
Here the luminosity is based on the distance from Baade-Wesselink methods. 
Carney \ea (1992) found masses as low as 0.45 \msun\ for 
low metallicity ([M/H] $\simeq -1.8$ dex) field RR~Lyr stars, 
but $M \simeq 0.55$ \msun\ for larger [M/H], 
still all below the theoretical expectation. 
 
In a first study using Hipparcos data, de Boer \ea (1997a) tried to 
resolve the $T_{\rm eff}$, $\log g$ and mass problem of the HB stars. 
Using the 8 best Hipparcos observed field HBA and sdB stars 
de Boer \ea (1997a) showed that they apparently have 
the low mass of $\simeq 0.4$ \msun\ as well. 
The similarity of the mass of the cluster HB stars and the field HB stars 
can be seen in Fig.\,2 of de Boer \ea (1997b). 

It has been claimed that there is a difference in $M_V$ between 
cluster RR Lyr stars and field RR Lyr stars (see Gratton 1998). 
Catelan (1998) doubts that such a difference exists and speculates 
that uncertainties in the basic stellar parameters and in the methods 
to derive them lie at the base of such claims. 

The absolute magnitude of the blue HB stars depends in a very sensitive way 
on the luminosity and the surface temperature. 
When going to hotter stars, $M_V$ will be fainter dramatically, 
leading to `drooping' HBs. 
Thus the blueest part of the HB is not useful to calibrate $M_V$ of the HB. 


\section{Kinematics indicates two populations of HB stars}

The studies of the kinematics of HB-like stars indicate 
that these stars form a rather inhomogeneous group. 

Using spectroscopic distances, radial velocities, and proper motions, 
de Boer \ea (1997c) showed that the sdB stars have, by and large, 
disk like orbits. 
As a sample the stars rotate along with the disk rather well 
with an asymmetric drift of only $-36$ \kms. 
The scale height of the sdB stars is of the order of 
300 pc (Aguilar Sanchez 1998). 

Altmann \ea (1998) investigated the kinematics of HBA stars. 
The orbits of several of these extend to large $z$. 
The asymmetric drift of their sample is nearly $-200$ \kms, 
indicating these stars are rather on halo like orbits. 

RR Lyrae stars as a group show the wide range of disk like to 
halo like orbits. 
The velocity in the direction of galactic rotation 
can be correlated with metallicity, giving the trend that 
metal rich RR Lyrae partake in the rotation of the disk, 
the metal poor stars (the majority) do not (Layden \ea 1996). 


The existence of {\it differences\,} in kinematic behaviour 
of RR Lyr and HBA stars on one hand and that of sdB stars on the other hand 
may indicate that the field HB star population differs significantly 
from the globular clusters. 
So is it allowed to assume that the field HB stars 
and the cluster HB stars are identical? 
Would it be possible to take the kinematic behaviour as an indication 
for [M/H] in the stars, and thus as indication for what $M_V$ ought to be?

\section{Brightness bias in data samples due to evolution and metallicity}

One has to take care of a brightness bias based on 
metallicity and on evolution when determining averages from samples. 

HB stars become brighter when evolving from the zero-age HB (ZAHB). 
Slowly the luminosity gets larger while the surface temperature 
stays almost the same. 
This means that HB stars cover a range in absolute magnitude. 
Inspection of the evolutionary tracks of Dorman (1992) 
shows the rise in $M_V$ amounts to about 0.07 mag. 
Thus, when studying a sample of field HB stars 
one is not dealing with a sample of the same $M_V$ for a given $B-V$. 
A nice example of the signi\-ficance of this aspect can be found in the 
ultraviolet CMD and the derived $L$, $T_{\rm eff}$ diagram 
for M~13 (Parise \ea 1998). 

The metal content of the stars has an effect on $M_V$. 
Lambert \ea (1996), basing themselves on numerous previous studies, 
arrive at a metal dependence of 
$M_V = 0.93 + 0.17 \times [{\rm Fe/H}]$ for RR Lyr stars 
(in which [Fe/H] really is only [M/H]). 
Thus $M_V$ may vary (for $-2< [{\rm Fe/H}] <0$) 
over $\simeq 0.34$ mag, also for field HB stars. 

The sum of the effects of evolution and of metallicity 
(assuming that the metallicity relation applies also to the 
star types adjacent to the RR Lyrae)
is that the metal poor evolved HB stars are the brightest, 
brighter by 0.4 mag than the metal rich ZAHB stars. 
The latter may thus be underrepresented in samples. 

\section{What do models tells us?}

The end of the RGB phase is marked by the ignition of He in the core 
of the star, leading to the transformation of the star into a ZAHB star. 
The He mass is then thought to be $\simeq 0.5$ \msun. 
The reddest HB stars have retained a rather heavy hydrogen shell and thus 
a non-negligible amount of hydrogen shell burning. 

Depending on the overall metallicity the H-burning contributes 
more or less to the overall luminosity. 
Here the details of CNO to Fe variations lead to considerable variations 
in luminosity (Dorman 1992). 
At $T_{\rm eff} = 10^4$~K, e.g., one finds (see his Fig.\,7) 
a 0.55 \msun\ star of [M/H]=$-0.47$ dex to have $\log L = 1.48$ 
while for a 0.68 \msun\ star existing at the same temperature with 
[M/H]=$-2.26$ dex the luminosity is $\log L = 1.58$. 

The core luminosity itself depends on the total mass of the star, 
as well as on the He-core mass. 
Rotation of the RGB star may lead to higher core masses 
(Mengel \& Gross 1976) with later a larger HB star core luminosity. 
However, this does not directly translate into a larger overall luminosity 
since the consequent larger mass loss at the RGB tip 
will lead to a lower H-shell luminosity. 
Modelling the effect of a larger He core mass 
(Caloi \ea 1997; Sweigart \& Catelan 1998) 
one finds for metal poor clusters brighter and bluer ends of the HB. 

Diffused He will alter the structure of the HB stars and the shape 
of the horizontal branch. 
Metal rich RHB stars with enhanced He will evolve away from the ZAHB 
into very extended blue loops, thereby producing a HB sloping up to 
brighter stars toward the blue (Sweigart \& Catelan 1998). 
Here the HB may be as bright as $M_V = 0$ mag. 

At the cooler end ($\log T_{\rm eff} = 3.85$) Cassisi \ea (1997) 
investigated the brightness of the RHB and the RGB bump. 
The absolute magnitude of the ZAHB varies from 
$M_V = 0.51$ mag for [M/H]=$-2.04$ to $M_V = 0.86$ mag for [M/H]=$-0.57$. 
The $M_V$ of the RGB bump is then $-0.20$ and 1.12 mag, respectively. 
It shows that the brightness {\it difference} between ZAHB and RGB bump 
changes over more than 1 mag in this metallicity range (Cassisi \ea 1997). 
Such models make clear that, 
due to the possible confusion of RHB stars with clump stars, 
candidate {\it field\,} RHB stars are utterly useless to calibrate the HB.

\begin{table}
\begin{center}
\caption[]{Field Horizontal Branch stars and RR Lyr}
\begin{tabular}{lrlllrlcc}
\hline
Name  &  $V$  &  $B-V$ & $A_V$ & [Fe/H] & $\pi$ & 
   $\sigma _{\pi}$ & $d$ & $M_V$ $\pm$ $\Delta M_V$\\ 
\cline{6-7} 
   &  &  &   & & \multicolumn{2}{c}{(mas)} & (pc)\\
\hline
HD \ \,86986      &  7.99 &   +0.12 &
         0.09 & $-1.9$ & 3.78 & 0.95 & 265 & +0.79 $\pm$ 0.55\\
HD 109995     &  7.62 &   +0.04 &
         0.00 & $-1.8$ & 4.92 & 0.89 & 205 & +1.08 $\pm$ 0.40\\
HD 130095     &  8.13 &   +0.08 &
         0.31 & $-2.0$ & 5.91 & 1.08 & 170 & +1.68 $\pm$ 0.40\\
HD 139961     &  8.85 &   +0.10 &
         0.31 &     & 4.50 & 1.19 & 220 & +1.81 $\pm$ 0.60\\
HD 161817     &  6.96 &   +0.16 &
         0.06 & $-1.6$ & 5.81 & 0.65 & 170 & +0.72 $\pm$ 0.25\\
RR Lyr        &  7.66 &   +0.25 & 
         0.09 & $-1.4$ & 4.38 & 0.59 & 228 & +0.78 $\pm$ 0.29\\
\hline
\end{tabular}

\noindent
HB star data from de Boer \ea (1997a), RR Lyr from Fernley \ea (1998)\\
$\Delta M_V$ given is due to the uncertainty in the parallax only
\end{center}
\end{table}

\begin{figure}
\begin{center}
\epsfig{file=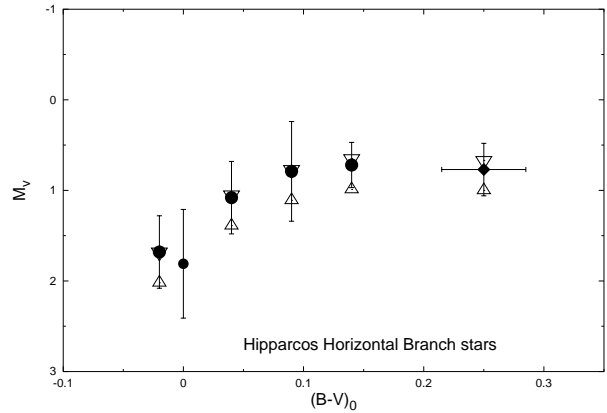,width=8cm}
\caption[]{The HB derived from stars with good Hipparcos parallaxes 
($\pi/\sigma_{\pi}\gtsim 4$)is presented. 
Absolute magnitudes for HB stars are given for the prototype HB's 
($\bullet$) from de Boer \ea (1997a). 
RR Lyr (filled $\diamond$), the only RR Lyrae star with good parallax 
(Fernley \ea 1998), is plotted at $B-V = 0.25$ with an errorbar in the colour. 
Using the $M_V$ versus [M/H] relation for RR Lyr stars, 
the HB for solar metallicty ($\Delta$) and for [M/H] = $-2$ ($\nabla$) 
is indicated
}
\end{center}
\end{figure}

\section{What does Hipparcos tell us about the HB?}

With Hipparcos about a dozen blue HB stars were measured of which 
only few have parallaxes accurate enough to be used to determine 
$M_V$ for the individual stars. 
These are essentially only the long known prototype HBA stars 
whose $M_V$ were presented by de Boer \ea (1997a). 
Their data ($\pi/\sigma_{\pi} \gtsim 4$) are given in Table 1. 

Several stars with colour and brightness like RHB stars 
are present in the Hipparcos data base. 
Given the risks of confusion with stars in different states of evolution 
(see Sect.\,5) we will ignore these objects. 

Stars of the blue HB, the red HB and RR Lyrae type have been used 
in an analysis by Gratton (1998). 
Many of these stars have less accurate parallaxes and thus deteriorate 
the quality of the ultimate averaged $M_V$. 
He assumes the HB should have a shape like that of M\,5 
(does M\,5 have the same history as the field stars?) 
and looks at the brightness differences between the {\it field\,} 
and the cluster HB shape. 
Then, his error weighting procedure is flawed (Popowski \& Gould 1998), 
and the analysis should have resulted in an absolute magnitude 
$\simeq 0.1$ mag fainter. 

A large number of RR Lyr type stars have been observed with Hipparcos. 
Only RR Lyr has a parallax large enough ($\pi/\sigma_{\pi} \gtsim 4$) 
to be included in Table 1. 
The absolute magnitude for the RR Lyraes as a group is not based on parallaxes 
but on the detected proper motions together with a model for their kinematics 
(Fernley \ea 1998). 
Since RR Lyrae stars have a range of metallicity and other intrinsic 
parameters, it is not proper to treat them as a single kinematic group. 
Thus, only RR Lyr is of use. 

Clearly, working directly with the best parallaxes (Table 1) 
provides the least ambiguous result. 
Plotting these best $M_V$ values 
one can {\it see\,} the HB for the field stars (Fig.\,1). 

An extrapolation to the blue edge of the RR Lyr strip, 
taking the plotted data at face value, 
gives $M_V \simeq 0.75$ mag (Seggewiss 1998). 

Assuming that the [M/H] values for these protoype HBA's are accurate, 
the $M_V$'s can be translated 
(using the relation found for RR Lyr stars, see Sect.\,8) 
into the solar metallicity HB. 
The observed stars so indicate that 
for [M/H]=0 and $(B-V)_0 = 0.20$ the $M_V \simeq 1.00$ mag, 
and for [M/H]=$-2$ (at $(B-V)_0 = 0.20$) that $M_V \simeq 0.65$ mag 
(see Fig.\,1).

\section{Consequences for studies of field HB stars}

From all of the above one must conclude that the assumption 
to be able to find one ultimate $M_V$ for HB-like stars 
is most likely wrong. 

\noindent
$\bullet$ HB stars in the {\it field\,} 
form a mix of old and metal poor stars with young 
and normal metallicity stars, thus intrinsically form a very 
inhomogeneous group. 
This is supported by the kinematic data for HB stars. 
The absolute magnitude of a star must therefore be normalised to 
a reference [M/H]. 
But, is [M/H] in the atmospheres of the observed stars the same as 
the original [M/H] (diffusion, levitation, convection)? 
Is then surface metallicity really correlated with $M_V$?
Or can one assume that all stars which kinematically are disk stars 
have normal [M/H] and that halo orbit stars have low [M/H]? 

\noindent
$\bullet$ For most stars the listed [Fe/H] values are based on 
a photometric index. 
This means that substantial uncertainties are present 
in the value for individual stars. 
For all the complexities see Layden (1994) and Lambert \ea (1996). 

\noindent
$\bullet$ The luminosity of the stars depends on the total mass 
and on the core mass. 
If the masses are different (smaller/larger) than the canonical 
$\simeq 0.5$ \msun\ 
then also the $M_V$ must be different (fainter/brighter). 

\noindent
$\bullet$ Observations indicate the existence of globular cluster 
horizontal branches slooping down as well as up toward the blue. 
Variations of He abundance may explain all these slopes. 
The differences in slope underline the possibility of large variations 
in $M_V$ for field HB stars. 

\noindent
$\bullet$ The nature of the gaps on the HB points to possibly different 
evolutionary routes and thus to possibily differences in $M_V$. 

\noindent
$\bullet$ Very blue HB stars have a large and very temperature sensitive 
range in $M_V$; they are not useable. 

\noindent
$\bullet$ Red HB stars are better to be avoided for general calibration work.

\section{Concluding remarks}

After the first paper dealing with $M_V$ for {\it field\,} HB stars 
based on Hipparcos parallaxes (de Boer \ea 1997a) numerous investigations 
came to different conclusions about the absolute magnitude of the HB. 
My suspicion is that all of the above discussed aspects of the 
intrinsic properties of the HB stars lie at the base of the discrepancies. 
Unfortunately, 
only {\it better parallaxes\,} and for more stars, 
such as hopefully will be obtained in the planned missions 
DIVA (R\"oser \ea 1997) and GAIA (Perrymann \ea 1997), 
can resolve the problems without ambiguity. 

\vspace{0.5cm}
\noindent
{\footnotesize 
I thank Wilhelm Seggewiss, Floor van Leeuwen, Michael Geffert 
and Martin Altmann for enlightening discussions.}

\vspace*{1cm}

\noindent
This text is the invited contribution to the conference on 13-16 Sept.\,1998 
in Haguenau, France:\\

``Harmonizing the Cosmic Distance Scale in the Post Hipparcos Era'',\\

D.\,Egret \& A.\,Heck (eds.), to be published in APS Conf.\,Ser.

\end{document}